# Buffer Management Algorithm Design and Implementation Based on Network Processors

Yechang Fang, Kang Yen
Dept. of Electrical and Computer Engineering
Florida International University
Miami, USA
{yfang003, yenk}@fiu.edu

Deng Pan, Zhuo Sun
School of Computing and Information Sciences
Florida International University
Miami, USA
{pand, zsun003}@fiu.edu

*Abstract*—To solve the parameter sensitive issue of the traditional RED (random early detection) algorithm, an adaptive buffer management algorithm called PAFD (packet adaptive fair dropping) is proposed. This algorithm supports DiffServ (differentiated services) model of QoS (quality of service). In this algorithm, both of fairness and throughput are considered. The smooth buffer occupancy rate function is adopted to adjust the parameters. By implementing buffer management and packet scheduling on Intel IXP2400, the viability of QoS mechanisms on NPs (network processors) is verified. The simulation shows that the PAFD smoothes the flow curve, and achieves better balance between fairness and network throughput. It also demonstrates that this algorithm meets the requirements of fast data packet processing, and the hardware resource utilization of NPs is higher.

I. INTRODUCTION

Network information is transmitted in the form of data flow, which constitutes of data packets. Therefore, different QoS means different treatment of data flow. This treatment involves assignment of different priority to data packets. Queue is actually a storage area to store IP packets with priority level inside routers or switches. Queue management algorithm is a particular calculation method to determine the order of sending data packets stored in the queue. Then the fundamental requirement is to provide better and timely services for high priority packets [1]. The NP is a dedicated processing chip to run on high speed networks, and to achieve rapid processing of packets.

Queue management plays a significant role in the control of network transmission. It is the core mechanism to control network QoS, and also the key method to solve the network congestion problem. Queue management consists of buffer management and packet scheduling. Generally the buffer management is applied at the front of a queue and cooperates with the packet scheduling to complete the queue operation [2, 3]. When a packet arrives at the front of a queue, the buffer management decides whether to allow the packet coming into the buffer queue. From another point of view, the buffer management determines whether to drop the packet or not, so it is also known as dropping control.

The control schemes of the buffer management can be analyzed from two levels, data flow and data packet. In the data stream level and viewed form the aspect of system resource management, the buffer management needs to adopt certain resource management schemes to make a fair and effective allocation of queue buffer resources among flows through the network nodes. In the data packet level and viewed from the aspect of packet dropping control, the buffer management needs to adopt certain drop control schemes to decide that under what kind of circumstances a packet should be dropped, and which packet will be dropped. Considering congestion control response in an end-to-end system, the transient effects for dropping different packets may vary greatly. However, statistics of the long-term operation results indicates that the transient effect gap is minimal, and this gap can be negligible in majority of cases. In some specific circumstances, the completely shared resource management scheme can cooperate with drop schemes such as tail-drop and head-drop to reach effective control. However, in most cases, interaction between these two schemes is very large. So the design of buffer management algorithms should consider both of the two schemes to obtain better control effects [4, 5].

## II. Existing Buffer Management Algorithms

Reference [6] proposed the RED algorithm for active queue management (AQM) mechanism [7] and then standardized as a recommendation from IETF [8]. It introduces congestion control to the router's queue operations. RED uses early random drop scheme to smooth packet dropping in time. This algorithm can effectively reduce and even avoid the congestion in network, and also solve the TCP protocol global synchronization problem.

However, one concern of the RED algorithm is the stability problem, i.e., the performance of the algorithm is very sensitive to the control parameters and changes in network traffic load. During heavy flow circumstances, the performance of RED will drop drastically. Since RED algorithm is based on best-effort service model, which does not consider different levels of services and different user flows, it cannot provide fairness. In order to improve the fairness and stability, several improved algorithms have been developed, including WRED, SRED, Adaptive-RED, FRED, RED with In/Out (RIO) [9, 10] etc. But these algorithms still have a lot of problems. For example, a large number of studies have shown that it is difficult to find a RIO parameter setting suitable for various and changing network conditions.

## III. The PAFD Algorithm

In this paper, we propose a new buffer management algorithm called PAFD (Packet Adaptive Fair Dropping). This algorithm will adaptively gain balance between congestion and fairness according to cache congestion situation. When there is minor congestion, the algorithm will tend to fairly drop packets in order to ensure all users access the system resources to their scale. For moderate congestion, the algorithm will incline to drop the packet of low quality service flows by reducing its sending rate using scheduling algorithm to alleviate congestion. In severe congestion, the algorithm will tend to fairly drop packets, through the upper flow control mechanism to meet the QoS requirements, and reduces sending rate of most service flows, in order to speed up the process of easing the congestion.

In buffer management or packet scheduling algorithms, it will improve the system performance to have service flows with better transmission conditions reserved in advance. But this operation will make system resources such as buffer space and bandwidth be unfairly distributed, so that QoS of service flows with poor transmission conditions cannot be guaranteed. Packet scheduling algorithms usually use generalized processor sharing (GPS) as a comparative model of fairness. During the process of realization of packet scheduling algorithms based on GPS, each service flow has been assigned a static weight to show their QoS. The weight $\phi_i$ actually express the percentage of the service flow $i$ in the entire bandwidth $B$. $\phi_i$ will not change with packet scheduling algorithms, and meet

$$\sum_{i=1}^{N} \phi_i = 1 \quad (1)$$

where $N$ expresses the number of service flows in the link. And the service volume is described by

$$g_i^{inc} = \frac{\phi_i}{\sum_{j \in B} \phi_j} B \quad (2)$$

where $i$, $j$ denotes two different service flows. In GPS based algorithms, the bandwidth allocation of different service flows meets the requirement $B_i/\phi_i = B_j/\phi_j$, where $B_i$ is the allocated bandwidth of the service flow $i$. By assigning a smaller weight $\phi$ to an unimportant background service flow, the weight of service flow with high priority $\phi_{high}$ will be much larger than $\phi_{low}$, so that the majority of the bandwidth is accessed by high-priority service flows.

### A. Algorithm Description

In buffer management algorithms, how to control the buffer space occupation is very key [11]. Here we define

$$\frac{C_i}{W_i} = \frac{C_j}{W_j} \quad (3)$$

where $C_i$ is the buffer space occupation, and $W_i$ expresses the synthetic weight of the service flow $i$. When the cache is full, the service flow with the largest value of $C_i/W_i$ will be dropped in order to guarantee fairness. Here the fairness is reflected in packets with different queue length [12, 13]. Assume that $u_i$ is the weight, and $v_i$ is the current queue length of the service flow $i$. The synthetic weight $W_i$ can be calculated as described by

$$W_i = \alpha \times u_i + (1-\alpha) \times v_i \quad (4)$$

where $\alpha$ is the adjust parameter of the two weighting coefficients $u_i$ and $v_i$. $\alpha$ can be pre-assigned, or determined in accordance with usage of the cache. $u_i$ is related to the service flow itself, and different service flows are assigned with different weight values. As long as the service flow is active, this factor will remain unchanged. $v_i$ is time varying, which reflects dropping situation of the current service flow.

Suppose a new packet $T$ arrives, then the PAFD algorithm process is described as follows:

- Step 1: Check whether the remaining cache space can accommodate the packet $T$, if the remaining space is more than or equal to the length of $T$, add $T$ into the cache queue. Otherwise, drop some packets from the cache to free enough storage space. The decision on which packet will be dropped is given in the following steps.

- Step 2: Calculate the weighting coefficients $u$ and $v$ for each service flow, and the parameter α. Then get the values of new synthetic weights $W$ for each flow according to (4).

- Step 3: Choose the service flow with the largest weighted buffer space occupation ($C_i/W_i$), if the service flow associated to the packet $T$ has the same value as it, then drop $T$ at the probability $P$ and returns. Otherwise, drop the head packet of the service flow with the largest weighted buffer space occupation at probability 1−$P$, and add $T$ into the cache queue. Here Probability $P$ is a random number generated by the system to ensure the smoothness and stability of the process.

- Step 4: Check whether the remaining space can accommodate another new packet, if the answer is yes, the packet will be transmitted into the cache. Otherwise, return to Step 3 to continuously choose and drop packets until there is sufficient space.

If all packet lengths are the same, the algorithm only needs one cycle to compare and select the service flow with the largest weighted buffer space occupation. Therefore, the time complexity of the algorithm is $O(N)$. In this case, we also need additional 4$N$ storage space to store the weights. Taking into account the limited capacity of wireless network, $N$ is usually less than 100. So in general the algorithm's overhead on time and space complexity are not large. On the other hand, if packet lengths are different, then it is necessary to cycle Step 3 and Step 4 until the cache has enough space to accommodate the new packet. The largest cycling times is related to the ratio between the longest and the shortest packets. At this moment, the time complexity overhead is still small based on practices.

In Step 2, α, a function of shared buffer, is a parameter for adjusting proportion of the two weighting coefficients $u$ and $v$. For a large value of α, the PAFD algorithm will tend to fairly select and drop packets according to the synthetic weight $W$. Otherwise, the algorithm tends to select and drop the service flow with large queue length. A reasonable value for α can be used to balance between fairness and performance. Here we introduce an adaptive method to determine the value of α. This adaptive method will determine α value based on the congestion situation of the cache, and this process does not require manual intervention.

When there is a minor congestion, the congestion can be relieved by reducing the sending rate of a small number of service flows. The number of service flows in wireless network nodes is not as many as that in the wired network. So the minor congestion can be relieved by reducing the sending rate of any one of service flows. We hope this choice is fair, to ensure that all user access to the system resources according to their weights.

When there is a moderate congestion, the congestion can not be relieved by reducing the sending rate of any one of service flows. Reducing the rate of different service flows will produce different results. We hope to reduce the rate of service flows which are most effective to the relief of congestion. That is, the service flow which current queue length is the longest (The time that these service flow occupied the cache is also the longest). This not only improves system throughput, but also made to speeds up the congestion relief.

When there is a severe congestion, it is obvious that reducing the sending rate of a small portion of the service flows cannot achieve the congestion relief. We may need to reduce the rate of a lot of service flows. Since the TCP has a characteristic of additive increase multiplicative decrease (AIMD), continuous drop packets from one service flow to reduce the sending rate would adversely affect the performance of the TCP flow. While the effect on relieving system congestion will become smaller, we gradually increase the values of parameters, and the algorithm will choose service flows to drop packet fairly. On one hand, at this point the "fairness" can bring the same benefits as in the minor congestion system; on the other hand this is to avoid continuously dropping the longer queue service flow.

Congestion is measured by the system buffer space occupation rate. $\alpha$ is a parameter relevant to system congestion status and its value is between 0 to 1. Assume that the current buffer space occupation rate is denoted by $Buffer_{cur}$, and $Buffer_{medium}$, $Buffer_{min}$, and $Buffer_{max}$ represent threshold value of the buffer space occupation rate for moderate, minor, and severe congestion, respectively.

When $Buffer_{cur}$ is close to $Buffer_{min}$, the system enters a state of minor congestion. When $Buffer_{cur}$ reaches $Buffer_{max}$, the system is in a state of severe congestion. $Buffer_{medium}$ means moderate congestion. If we value $\alpha$ by using linear approach, the system will have a dramatic oscillation. Instead we use high order nonlinear or index reduction to get smooth curve of $\alpha$ as shown in Figure 1.

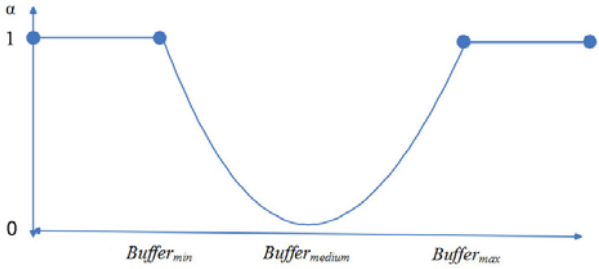

Fig.1. An adaptive curve of α

The value of $\alpha$ can also be calculated as below

$$\alpha = \begin{cases} 0, & \text{if } Buffer_{cur}^2 < Buffer_{min}^2 \\ 1 - \dfrac{Buffer_{cur}^2 - Buffer_{min}^2}{Buffer_{max}^2 - Buffer_{min}^2}, & \text{if } Buffer_{min}^2 \leq Buffer_{cur}^2 \leq Buffer_{max}^2 \\ 1, & \text{if } Buffer_{cur}^2 < Buffer_{max}^2 \end{cases} \quad (5)$$

### B. DiffServ Model Support

In the PAFD algorithm, we can adopt the DiffServ model to simplify the service flows by dividing them into high-priority services such as assurance services and low-priority services such as best-effort services. We use the queuing method for the shared cache to set and manage the cache. When a new packet arrives at the cache, first the service flow is checked to see whether it matches the service level agreement (SLA). If it does, then this new packet enters the corresponding queue. Otherwise, the packet is assigned to low-priority services, and then enters the low-priority queue.

In the DiffServ model, we retain the implement process of PAFD, and only modify (4) into

$$W_i = (\alpha \times u_i + (1-\alpha) \times v_i) \times \beta \quad (6)$$

where $\beta$ is a new parameter used to adjust the fairness among service flows of different service levels. As mentioned above, we can set the value of parameter $\alpha$ different from that shown in Figure 1 to satisfy different requirements. $\alpha$ is the parameter which balances fairness and transmission conditions. For high-priority services, the curve in Figure 1 is reasonable. The fairness is able to guarantee the QoS for different service flows, and also is required to relief congestion quickly. For high-priority services which have no delay constraints and high fairness requirements, a higher throughput is more practical. Therefore, we can get the value of the parameter α for low-priority services, which is slightly less than that for high-priority services as shown in Figure 2.

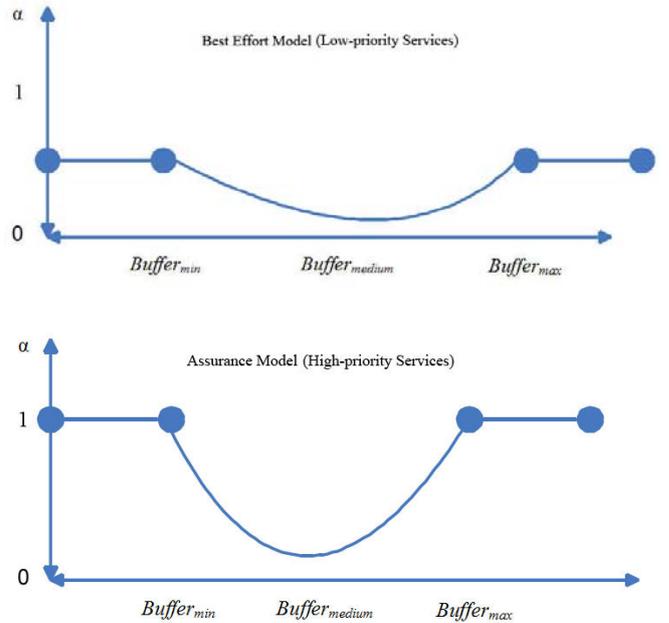

Fig.2. Values of α for different priority services

Now we check the effects of the parameter $\beta$. For high-priority services, $\beta$ is a constant with value 1. For low-priority services, the value of $\beta$ is less than 1, and influenced by the network load. When network load is low, $\beta$ equals to 1. In this case, different level service flows have the same priority to share the network resources. As network load increases, in order to guarantee the QoS of high-priority services, low-priority services gradually give up some transmission opportunities, so the value of $\beta$ decreases. The higher network load is, the smaller the values of *β* and *W* are. Therefore, the probability of a low-priority packet being dropped is higher. Values of *β* are shown below.

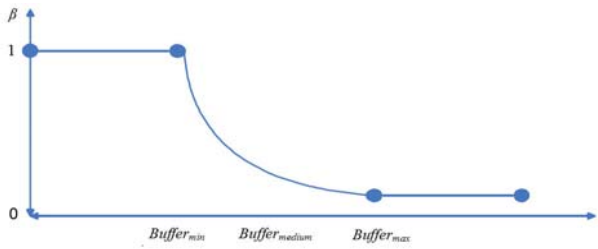

Fig.3. Values of β for different priority services

## IV. SIMULATION RESULTS

### A. Simulation for Commen Services

We compare the PAFD algorithm with two commonly used buffer management algorithms RED and tail drop (TD). We choose two common packet scheduling algorithms Best Channel First (BCF) and Longest Queue First (LQF) to work with PAFD, RED and TD. Here the LQF uses the weighted queue length for packet scheduling. So there are 6 queue management algorithm combinations, which are PAFD-BCF, PAFD-LQF, RED-BCF, RED-LQF, TD-BCF, and TD-LQF. The performance comparisons of these algorithms are carried out with respect to throughput effectiveness, average queuing delay, and fairness.

We use K1297-G20 signaling analyzer to simulate packet sending, and the operation system for K1297-G20 is Windows NT 4.0. ADLINK 6240 is used as the NP blade. Based on the simulation configuration, there are 8 different packet length configurations for the data source. They are fixed length of 64 bytes, fixed length of 65 bytes, fixed length of 128 byte, fixed length of 129 bytes, fixed length of 256 bytes, random length of 64-128 bytes, random length of 64-256 bytes, and random length of 64-1500 bytes.

Figure 4 shows that all the algorithms have similar throughputs for low network load. When the load increases, the throughput effectiveness of BCF is higher than that of other scheduling algorithms. This figure shows that PAFD-BCF provides significant higher throughput than the other algorithms. PAFD does not randomly drop or simply tail drop packets, but fully considers fairness and transmission conditions. In this way, service flows under poor transmission condition receive high probability of packet dropping, thus a relatively short virtual queue. When BCF is working with PAFD, the service flow under better channel transmission condition will give higher priority and result effective throughput.

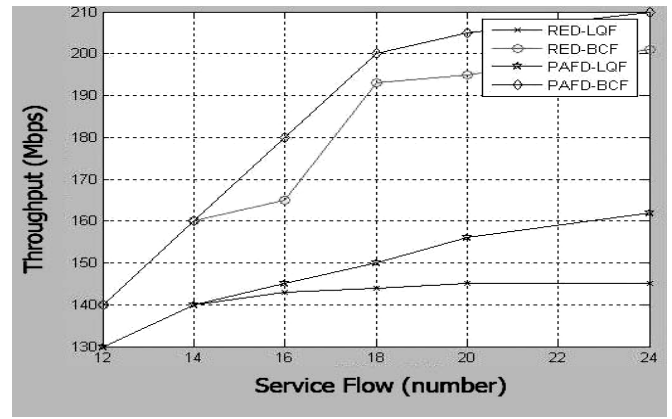

Fig.4. Throughputs of RED and PAFD

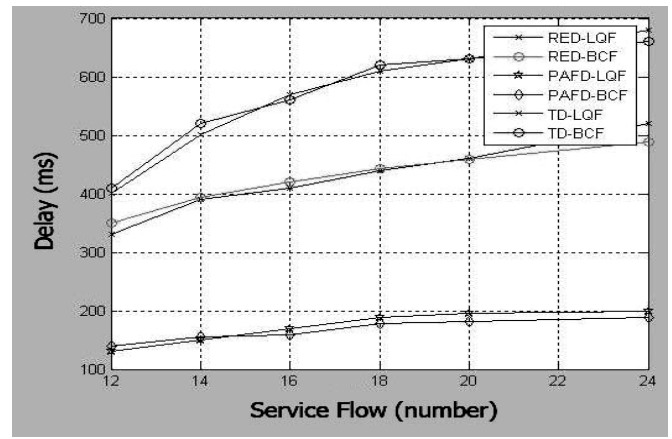

Fig.5. Average Queuing Delay for TD, RED and PAFD

From Figure 5, we find that RED has better performance on the average queuing delay due to the capability of early detection of congestion and its drop mechanism. BCF has better performance on queuing delay than that of LQF. As the load increases, the average queuing delay of PAFD first increases, then decreases. This is because RAFD does not use tail drop, and instead searches a service flow with the largest weighted buffer space occupation to drop the head packet to reduce the average queuing time.

Both TD and RED use shared cache instead of flow queuing so that they fail to consider the fairness. Here the fairness index *F* is given by

$$F = \frac{(\sum_{1}^{N} \frac{G_i}{W_i})^2}{N \sum_{1}^{N} (\frac{G_i}{W_i})^2} \quad (7)$$

where $G_i$ is the effective throughput of service flow *i*, and *N*

is the total number of service flows. It is not difficult to prove that $F \in (0, 1)$. When $F$ has a lager value, the fairness of the system is better. If the value of $F$ equals to 1, the system resource is completely fair. We can use (7) to calculate the fairness index and compare the fairness of different algorithms. In ON-OFF model with the assumption that there are 16 service flows, the ON average rate of flows 1-8 is twice of that of 9-16. That is, $W_i : W_j = 2 : 1$, where $i \in [1, 8]$ and $j \in [9, 16]$. Using round robin algorithms without considering $W$, we can calculate the reference value of fairness index $F = 0.9$. Table I gives the fairness index of TD, RED and PAFD which are combined with packet scheduling algorithms.

TABLE I. FAIRNESS INDEX

| Algorithms | Fairness |
|---|---|
| TD-BCF | 0.8216 |
| TD-LQF | 0.9162 |
| RED-BCF | 0.8855 |
| RED-LQF | 0.9982 |
| PAFD-LQF | 0.9988 |
| PAFD-BCF | 0.8902 |

The table indicates that the fairness index of BCF is lower when combined with TD and RED. Since PAFD takes the fairness into consideration, the fairness index of PAFD is higher than that of TD when there are congestions. The combination of PAFD and LQF has higher throughput and more fair distribution of cache and bandwidth resources. By changing the value of parameter $\alpha$, we can conveniently balance the system performance and fairness based on the requirements.

### B. Simulation for DiffServ Model

In this section we adopt the same environment as described in the previous section to test the PAFD performance based on the DiffServ model. The only difference is that half of the services are set to high-priority, and another half to low-priority.

Figures 6 and 7 show the throughput and average queuing delay of those algorithms. The only difference in these two tests is that the value of parameter $\alpha$ for half of the service flows used in the second simulation is slightly lower than the one in the first simulation. So the curves in Figures 7 and 8 are very similar to those shown in Figures 4 and 5.

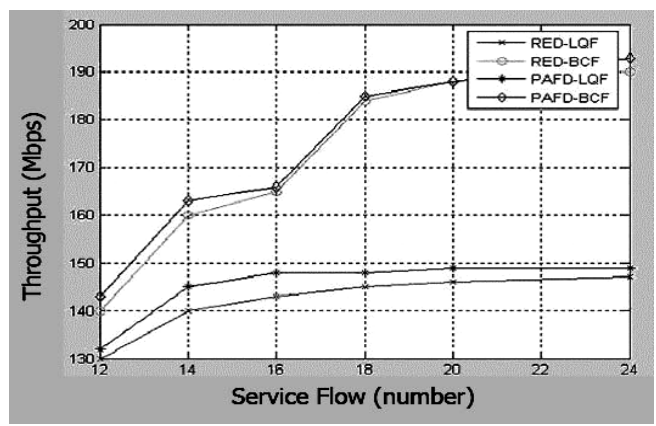

Fig.6. Throughputs of RED and DS-PAFD

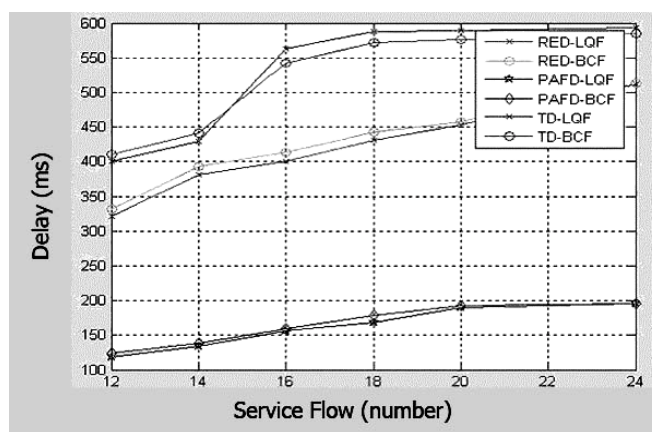

Fig.7. Average Queuing Delay of RED and DS-PAFD

Table II gives the comparison of fairness index of theses algorithms. Comparing these numbers with those shown in Table I, we can draw a similar conclusion. However, the difference in values is that the fairness index of low-priority services is slightly lower than that of high-priority services as a result of different values of parameter $\alpha$ selected.

TABLE II. COMPARISON OF FAIRNESS INDEX

|  | TD-BCF | TD-LQF |
|---|---|---|
| Flow | 0.8346 | 0.9266 |
|  | DSPAFD-BCF | DSPAFD-LQF |
| High-priority Service Flow | 0.8800 | 0.9922 |
|  | DSPAFD-BCF | DSPAFD-LQF |
| Low-priority Service Flow | 0.8332 | 0.9488 |

As shown in Figures 2-3, 6 and 7, when network load is light, the throughputs are similar for different priority services. This means different priority services have the same priority to share network resources. As network load increases, the throughput gradually decreases. However, even in the case of heavy load, the PAFD algorithm still allocates small portion of resources to low-priority services to meet the fairness requirement. And this operation will prevent high-priority services from fully occupying the network resources.

## V. IMPLEMENTATION BASED ON NETWORK PROCESSORS

Here we adopt NP Intel IXP2400 to implement the PAFD algorithm. Intel IXP2400 provides us with eight micro-engines, and each micro-engine can support up to eight hardware threads. When the system is running, each micro-engine deals with one task. During the thread switching, there is no need for protection, each hardware thread has its own register, so the switching speed is very fast. Also Intel IXP2400 is appropriate for DiffServ model.

The PAFD Algorithm executes enqueuing and dequeuing operations in the transmission, which are implemented using chained list of the SRAM of IXP2400. The buffer manager of PAFD receives enqueuing request from the functional pipeline, and accepts dequeuing request through the micro engines of NPs. In the PAFD algorithm, Q-Array in the SRAM controller is used to the chained list, and a queue descriptor is stored in the SRAM. The buffer manager uses content associative memory (CAM) to maintain queue buffer of the descriptor. When enqueuing request arrives, the buffer manager will check CAM to see if the queue descriptor is in the local buffer. If so, PAFD will be run to decide whether the new packets should enter the queue. If not, the descriptor is excluded from the Q-Array, and then stored in the SRAM. Therefore, another specified queue descriptor is read into the Q-Array, and then PAFD is run to decide whether to drop the new packets. When a queue enters a queue, Q-Array logic moves the first four bits to the SRAM controller. Q-Array can buffer 64 queue descriptors in each SRAM channel. The PAFD algorithm only reserves 16 entrances for the buffer manager, and the rest are for free idle chained list and SRAM loops. The current count of packets is stored in the local memory. This operation needs 16 bits, and each bit represents the number of packets through the 16 entrances. The packet counter is initialed when entrances are read into the Q-Array, and then it executes the operation of plus one or minus one base on the response. The implemented system we designed supports 64 virtual ports, and each port supports 16 queues. Thus, there are 1024 queues in total. As we adopt the SRAM structure, it is very easy to enqueue.

The dequeuing operation is similar to the enqueuing operation. In order to maintain the performance of the system, micro engine threads of NPs must operate in strict accordance with the predetermined sequence. This is controlled by internal thread semaphore. When a queue changes from empty to non-empty in an enqueuing operation, or from non-empty to empty in a dequeuing operation, the buffer manager of PAFD will send a message to packet scheduling module through the adjacent loop.

## VI. CONCLUSIONS

Buffer management algorithm is the core mechanism to achieve network QoS control. It also plays an important role in network resource management. In this paper, a novel buffer management algorithm called PAFD is proposed based on NPs. The PAFD algorithm takes into account the impact of transmission environment on packets. It can adaptively balance between queue congestion and fairness according to cache congestion. PAFD also supports the DiffServ model to improve network QoS based on NPs. The simulation results show that the throughput and fairness are better balanced after this algorithm is applied. Finally, the PAFD algorithm is implemented based on IXP2400, which means that the hardware resource utilization of NPs is higher.

The future network has two development requirements: high-speed bandwidth and service diversification. Research on buffer management algorithms is able to suit for these requirements. In the future, buffer management will become more complex. Therefore, the requirements for NPs and other hardware will be more stringent. It is very important to consider the comprehensive performance of the algorithms while pursuing simplicity and easy implementation.

ACKNOWLEDGEMENTS

This work was supported by Presidential Fellowship 2007-2009 and Dissertation Year Fellowship 2009-2010, Florida International University.

## AUTHORS PROFILE

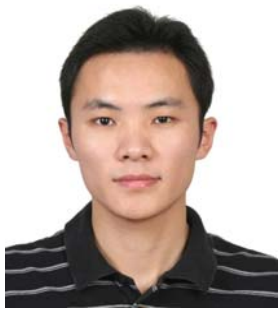

**Yechang Fang** received his M.S. in Electrical Engineering from Florida International University (FIU), Miami, USA in 2007. From 2006 to 2007, he served as an IT specialist at IBM China to work with Nokia, Motorola and Ericsson. He is currently a Ph.D. candidate with a Dissertation Year Fellowship in the Department of Electrical and Computer Engineering, FIU. His area of research is telecommunication. Besides, his research interests also include computer networking, network processors, fuzzy Logic, rough sets and classification.

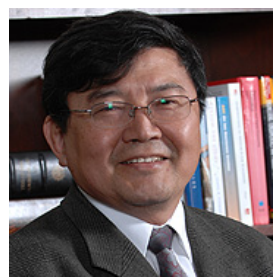

**Kang K. Yen** received the M.S. degree from University of Virginia in 1979 and Ph.D. degree from Vanderbilt University in 1985. He is currently a Professor and Chair of the Electrical Engineering Department, FIU. He is also a registered professional engineer in the State of Florida. He has been involved in theoretical works on control theory and on parallel simulation algorithms development for real-time applications in the past several years. In the same periods, he has also participated in several industry supported projects on real-time data processing and microprocessor-based control system designs. Currently, his research interests are in the security related issues and performance improvement of computer networks.

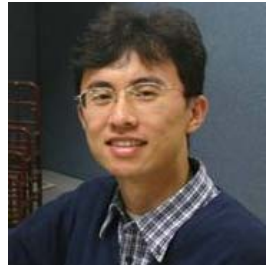

**Deng Pan** received his Ph.D. and M.S. degree in Computer Science from State University of New York at Stony Brook in 2007 and 2004. He received M.S. and B.S. in Computer Science from Xi'an Jiaotong University, China, in 2002 and 1999, respectively. He is currently an Assistant Professor in the School of Computing and Information Sciences, FIU. He was an Assistant Professor in School of Computing and Information Sciences, FIU from 2007 to 2008. His research interests include high performance routers and switches, high speed networking, quality of service, network processors and network security.

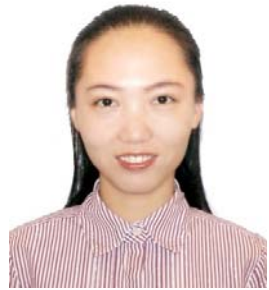

**Zhuo Sun** received her BS degree in computer science from Guangxi University, Nanning, China, in 2002, and the MS degree in software engineering from Sun Yat-sen University, Guangzhou, China, in 2005. Then she worked at Nortel Guangzhou R&D, Guangzhou, China. She is currently a second year Ph.D student in Florida International University. Her research interests are in the areas of high-speed network.